\documentclass{JHEP3}

\def\pa{\partial}
\def\na{\nabla}
\def\nn{\nonumber}

\newcommand{\C}{\mathcal{C}}
\newcommand{\LGB}{\mathcal{L}_\mathrm{GB}}
\newcommand{\be}{\begin{equation}}
\newcommand{\ee}{\end{equation}}
\newcommand{\bea}{\begin{eqnarray}}
\newcommand{\eea}{\end{eqnarray}}
\newcommand{\bref}[1]{(\ref{#1})}

\title{Scalar brane backgrounds in higher order curvature gravity}
\author{Christos Charmousis, Stephen C. Davis\thanks{New address: 
Institut de physique th\'eorique, BSP, EPFL, 
CH--1015 Lausanne, Switzerland} \ and Jean-Fran\c{c}ois Dufaux \\ 
LPT, Universit\'e de Paris-Sud, B\^at. 210, 91405 Orsay CEDEX, France \\
E-mail: \email{Christos.Charmousis@th.u-psud.fr},
\email{Stephen.Davis@th.u-psud.fr} and \email{Dufaux@th.u-psud.fr}}

\abstract{We investigate maximally symmetric brane world solutions with
a scalar field. Five-dimensional bulk gravity is described by a 
general Lagrangian which yields field equations containing
 no higher than second
order derivatives. This includes the Gauss-Bonnet combination for the
graviton. Stability and gravitational properties of such solutions are
considered, and we particularily emphasise the modifications  induced by the
higher order terms. In particular it is shown that higher curvature
corrections to Einstein theory can give rise to instabilities in
brane world
solutions. A method for analytically obtaining the general solution for
such actions is outlined. Generically,
the requirement of a finite volume element together with the 
absence of a naked
singularity in the bulk imposes fine-tuning of the brane tension. A
model with a moduli scalar field is analysed in detail and we address
questions of instability and non-singular self-tuning solutions. 
In particular, we discuss a case 
with a normalisable zero mode but infinite volume
element.}

\keywords{eld, ctg}
\preprint{LPT-Orsay-0358}

\begin{document}

\section{Introduction}

Developments in string theory suggest that matter and gauge interactions
(described by open strings) may be localised on a brane, embedded into a
higher dimensional spacetime. Fields represented by closed strings, in
particular gravity, propagate in the whole of spacetime. Such an idea
provides an interesting framework to address theoretical issues of
particle physics and cosmology, such as the hierarchy and cosmological
constant problems, or the source of dark energy and dark matter. Several
toy models were introduced (see for example~\cite{lukas, BDL, early} for
early works) to address some of these problems. Some directly emerged
from string theory setups \cite{horava} while others were merely
inspired by certain aspects of string theory. The possibility of large
extra dimensions was considered in ref.~\cite{ADD} for an unwarped
spacetime. It was later shown~\cite{RS} that warping of five-dimensional
spacetime could lead to localisation of gravity on the brane even
though the extra dimension was of infinite proper length. Gravity in
such models seems four-dimensional down to very small distances where
higher (discrete or continuum) Kaluza-Klein modes take over. Later on
several models were proposed where gravity was actually quasi-localised,
i.e.\ four-dimensional at intermediate scales and modified at very large
distances (see for example~\cite{DGP, GRS, papa}). Such models yield
interesting possibilities for the origin of dark energy~\cite{deffayet},
`observed' in cosmological experiments. However such models seem
generically to suffer from ghosts~\cite{luigi} or strong coupling
problems~\cite{rubakov}.

Let us step back to the case of localised gravity. As we mentioned
above, gravity may be localised in the vicinity of a distributional,
positive-tension, flat four-dimensional brane, embedded in a
five-dimensional anti-de Sitter bulk, with an infinite extra
dimension. The presence of a non trivial background is essential
here. More precisely, the Einstein equations in this case admit a
solution of the form
\be
\label{poinc}
ds^2 = e^{2 A(z)}\,\eta_{\mu \nu}\,dx^{\mu}dx^{\nu} + dz^2 \ .
\ee
The tension of the brane needs to be fine-tuned relative to the bulk
cosmological constant in order to give four-dimensional Poincar\'e
invariance. This fine-tuning of parameters corresponds to the usual
cosmological constant problem in this context. Linear perturbations
around this background show the existence of a single massless
zero-mode, together with a continuum of massive Kaluza-Klein modes. The
zero-mode is normalisable (because the volume element $\int e^{2 A} dz$
is finite) and localised on the brane. This is interpreted as the usual
four-dimensional graviton. Furthermore, the wave-functions of the
massive modes are highly suppressed on the brane, so that their
contribution to four-dimensional gravity at low energy may be
negligible. This was further confirmed, at the linear level, by a
rigorous derivation of Newton's law experienced by observers on the
brane~\cite{GT}.

The Randall-Sundrum model~\cite{RS} and its basic features rely on
assumptions, such as the localisation of matter on a distributional
source (see however~\cite{dubovsky}), a constant curvature bulk, and
General Relativity to describe gravity in higher than 4 dimensions. One
would like to know how generic these assumptions are. Since gravity
plays such a key role in brane world models, it is certainly interesting
to investigate how the usual features are modified by more general
gravitational theories and backgrounds which are still compatible at
some limit with Einstein's theory and fundamental principles, and are
also in accord with string theory low energy effective actions. This is
the direction we take in this paper.

Toward this aim we add higher curvature contributions to the usual
Einstein-Hilbert and cosmological constant terms in the bulk action. In
four dimensions the Einstein-Hilbert term is the only term (apart from
the cosmological constant) which yields at most second order derivatives
in the field equations. In five dimensions this is no longer
true~\cite{lanczos}, and the quadratic Gauss-Bonnet term
\be
\LGB=R^2 - 4 R_{ab}R^{ab} + R^{abcd} R_{abcd} 
\label{LGB}
\ee
also has this property. Any other quadratic term will produce higher
order derivatives in the field equations, which will give rise to ghosts
in the theory. Thus the Gauss-Bonnet combination is the only other
sensible curvature term which can be included in the action for gravity
(in five dimensions).

The Gauss-Bonnet term is the Euler characteristic in four dimensions
just as the Ricci scalar is the Euler characteristic of two
dimensions. The similar properties of the two terms are direct
consequences of this fact. In an arbitrary number of dimensions the most
general action yielding second order field equations has been given by
Lovelock~\cite{lovelock} (see also~\cite{zumino} for an elegant
derivation using differential forms). Brane gravity with the
Gauss-Bonnet term in a constant curvature bulk spacetime was first
studied in \cite{kim}{\footnote{see however \cite{GBlin},
\cite{meissner} for the junction conditions}}.
A careful treatment of the graviton fluctuations was given in
 ref.~\cite{GBlin} (see also~\cite{meissner} for the full use
of Lovelock theory in arbitrary dimensions). Cosmological evolution and
consequences of such a setup have also been studied (see for 
example~\cite{fax, steph, gravanis, lidsey, 
Gregory:2003px, Kofinas}). For some early
applications to cosmology see \cite{mad}.

Another generalisation is to add a scalar field, with a Liouville type
potential in the bulk. This may represent the dilaton and/or moduli
fields of compactified string theory. In this context, gravitational
fluctuations around scalar brane backgrounds have been studied (see
e.g.~\cite{scagrav}, and also~\cite{CEHS} for a smooth brane
realisation). One of the phenomenological benefits of the inclusion of
such a scalar field is an interesting reformulation of the cosmological
constant problem in brane world models~\cite{ADKSS}. One is then able to
use the additional degrees of freedom to ensure the existence of a
four-dimensional Poincar\'e invariant solution whatever the values of
the brane tension and the bulk cosmological constant. Thus there is no
fine tuning of the fundamental parameters of the theory. This is the
so-called `self-tuning' mechanism. Unfortunately scalar field brane
worlds with Liouville potentials generically suffer from naked
singularities at finite proper distance from the brane, at least when
sensible four-dimensional gravity is required~\cite{sing}. This is true
in particular for the self-tuning solutions~\cite{CEGH}. The presence
of the singularities re-introduces fine-tuning into the problem. This
therefore represents a rephrasing, and not a solution, of the
cosmological constant problem~\cite{FLLN}.

Close to these naked singularities, the original effective gravitational
theory breaks down, because higher order terms in the Lagrangian cannot
be neglected anymore. We are thus naturally led to consider higher order
curvature terms and, for the same reasons, higher order scalar kinetic
terms as well. In this paper we will consider terms which are up to
quartic order in derivatives in the action. Such actions arise naturally
in the context of low energy effective string theory~\cite{Mets}, and
are known to lead to interesting properties in four-dimensional
cosmology~\cite{ART} as well as in black-hole physics~\cite{GBH}. Brane
worlds with actions of this kind have been previously studied for a
metrics of the form~(\ref{poinc}) with $A \propto z$. The basic setup
and particular solutions were analysed in detail in
ref.~\cite{mavromatos2}, and asymptotic behaviour of solutions and
self-tuning was discussed in ref.~\cite{Jakobek} (see also \cite{neu}). 
In ref.~\cite{us}, an
exact solution was found with non trivial scalar field profile, no naked
singularity and finite four-dimensional Planck mass. However, the
`self-tuning' mechanism was shown to fail since the solution was fine-tuned 
much in the same way as in the Randall-Sundrum model. In fact, as
explained in section~\ref{fa}, in order to address this latter issue,
\emph{general} solutions have to be known analytically. The
construction, analysis and stability study of such solutions is one of
the aims of this work.

The organisation of the paper is as follows: In the next section, we
discuss the general action that we will consider. We demonstrate in
particular that the origin of the scalar field (e.g.\ dilaton or moduli)
fixes the coefficients of the higher order terms. We give the
coefficients for toroidal Kaluza-Klein compactification. Field
equations and boundary conditions for a conformally flat brane
background~(\ref{poinc}) are discussed in section~\ref{fa}. Next, in
section~\ref{gm}, we study the spin 2 
linear gravitational perturbations around an arbitrary conformally flat
background. We point out in all generality 
the essential differences of spin 2 perturbations of
Einstein-Gauss-Bonnet  with respect to
ordinary Einstein gravity. We address in particular the question of gravity
localisation on the brane. It is also shown generically that stability
requirements are far more stringent in higher curvature brane world
setups. This essentially means that Einstein brane worlds can develop
order $\alpha'$ instabilities in an effective action approach. Having
discussed in a general context background and perturbations, we move on
to demonstrate how one finds solutions and their gravitational
properties (singularities, localisation of gravity, stability etc). In
section~\ref{cfs} we show how to obtain the general conformally flat
solution to the field equations and their basic properties. 
In section~\ref{KK6D} we apply our general analysis 
to the algebraically simplest case where the scalar field comes from the
compactification of a flat sixth dimension. It is shown that 
the general maximally
symmetric brane world solution is obtained from the corresponding
six-dimensional black-hole solution. We concentrate on the cases
without naked singularities in the bulk and study their stability and
self-tuning requirements. It is intriguing that in the sole case where
`self-tuning' is actually possible (normalisable graviton zero-mode
without naked singularity in the bulk, nor brane tension tuning), a
ghost appears in the bulk. The last section contains a summary of our
main results and our conclusions.

\section{Action}

\label{2}

Consider the action describing a scalar field which is conformally coupled to
gravity, written in the Jordan (or string) frame, with terms which are
up to quartic order in the field derivatives
\bea
\label{action}
S &=&\frac{M^{3}}{2}\int d^5x \sqrt{-g} \, e^{-2\phi}
\biggl\{R -4\omega \pa_a\phi\pa^a\phi-2\Lambda
\nonumber \\
&+& \alpha \left[\LGB+16a \pa_a\phi\pa^a\phi\; \Box\phi
-16b G^{ab} \pa_a\phi\pa_b\phi
-16c (\pa_a\phi\pa^a\phi)^2\right]\biggr\} \ .
\eea
This is the most general action of this order which obeys the rules of
Lovelock gravity. This essentially means that the field equations
obtained by variation of the scalar field $\phi$ and the metric tensor
$g_{\mu\nu}$ are linear in second order derivatives, and do not feature
higher order derivatives. This is an essential and natural hypothesis
when studying backgrounds with boundaries and linear perturbations of
them. In a string theory context the above frame is referred to as the
string frame if $\phi$ is the dilaton. Otherwise (for example if $\phi$
is a moduli field) it should be referred to as a Jordan frame.

The constants $a$, $b$ and $c$ express the higher order terms for the
scalar field and its interaction with the Einstein tensor. They will be
given specific values later on, depending on the physical nature of the
scalar field. $\omega$ is the Brans-Dicke parameter. The higher order
terms couple via the dimensionfull constant $\alpha$ which is of
dimension $(length)^2$. For string theory it is proportional to the
string tension $\alpha'$. Any terms other than those appearing in the above
action~\bref{action} yield higher than second order derivatives, which always
result in the existence of ghosts~\cite{zwei}, even around a flat
background.

Scalar-tensor theories of gravity are also studied in the Einstein
frame, which is related to the Jordan frame by the conformal transformation
$g_{ab}\rightarrow e^{-4\phi/3} g_{ab}$. The Einstein frame action is
\bea
\label{actionE}
S_{E}&=&\frac{M^{3}}{2}\int d^5x \sqrt{-g} \, 
\biggl\{R -\frac{4}{3}\,(3\omega+4) \pa_a\phi\pa^a\phi
-2\Lambda\,e^{4\phi/3}
\nonumber \\ 
&+& \alpha\,e^{-4\phi/3}\, \left[\LGB+16\tilde a\,\pa_a\phi\pa^a\phi\; \Box\phi
-16\tilde b\, G^{ab} \pa_a\phi\pa_b\phi
-16\tilde c\, (\pa_a\phi\pa^a\phi)^2\right]\biggr\} 
\eea
where
\be
\label{stoe}
\tilde a=a-3b+\frac{16}{9} \hspace*{0.3 cm} , \hspace*{0.3 cm}
\tilde b=b-\frac{16}{9} \hspace*{0.3 cm} , \hspace*{0.3 cm}
\tilde c=-2a+\frac{4}{3}b+c-\frac{8}{27} \hspace*{0.3 cm} .
\ee
The identifications between the coefficients are easily obtained by
comparing the field equations in the two frames (see Appendix). Written
in this form the action more closely resembles the usual
Einstein-Hilbert action.

If the scalar field $\phi$ plays the role of the dilaton, 
the parameters $a$, $b$ and $c$ in eq.~\bref{action}
are constrained in order to reproduce the scattering string amplitudes
on shell \cite{Mets}. For $\omega=-1$ this constraint reads,
\be
\label{mets}
2a-2b-c+1=0 \ .
\ee
The case $a=b=c=1$ exhibits additional symmetry (higher-order
extension of T-duality~\cite{M}, as well as $\sigma$-model conformal
invariance~\cite{MM}). In the context of brane world models, it has been
considered in ref.~\cite{Jakobek}, while the case $\tilde a=\tilde b=0$,
$\tilde c=1/27$ has been discussed in detail in refs.~\cite{mavromatos2,us}. 

Another interesting case arises from toroidal Kaluza-Klein 
compactification. Start by considering the $5+N$-dimensional action
\be
\label{caction}
S^{(5+N)}=\frac{M^3}{2L^N}\int d^5x dX^N {\textstyle\sqrt{-g_{(5+N)}}} 
\left\{R^{(5+N)} -2\Lambda + \alpha \LGB^{(5+N)}\right\} 
\ee
and take the $N$ extra dimensions to be flat and compact, with
$0 \leq X_A < L$ and $A,B=1 \ldots N$. Then for the metric ansatz
\be
\label{kk}
ds_{5+N}^2 = g_{ab}(x) dx^a dx^b 
+ e^{-4\phi(x)/N} \eta_{AB} dX^A dX^B
\ee
we obtain the same field equations as eq.~\bref{action} with 
\be
\label{kaka}
\omega = -\frac{N-1}{N} \ , \ \ a = (2\omega+1)\omega \ , \ \
b = -\omega \ , \ \ c = -(2\omega+1)\omega^2 \ .
\label{kkparam}
\ee
This is the extension to higher curvature gravity theories of the usual
Kaluza-Klein toroidal compactification. The scalar field in this case
arises simply as the unique size or moduli of the $N$ compact extra
dimensions. For instance, in the string frame, $N=1$ is simply
$\omega=a=b=c=0$, and $N=2$ is $\omega=-1/2$, $a=c=0$, $b=1/2$. As $N$
increases the extra terms in the action (\ref{action}) are switched on,
adding successive layers of complication. Interestingly the special case
$a=b=c=1$ of eq.~\bref{mets} corresponds to $N \rightarrow \infty$ and
$\omega=-1$.

\section{Field Equations}

\label{fa}

Consider the conformally flat metric,
\be
\label{ansatzB}
ds^2 = e^{2A(z)} dx^\mu dx^\nu \eta_{\mu \nu} + e^{2B(z)} dz^2 \ .
\ee
Four dimensional spacetime sections are Poincar\'e invariant, and 
the single extra dimension (with coordinate label $z$) is warped. 
Note that we still have a gauge freedom relating the metric components $A$,
$B$ and $\phi$; for example $B=0$ corresponds to a Gaussian normal
system. We shall use this gauge freedom to solve the field equations in
convenient coordinate systems in sections \ref{cfs} and \ref{KK6D}.
The full field equations are listed in the appendix. 
The (5,5) component of the generalised Einstein equation gives
\bea
\label{55} 
&&e^{2B}\Lambda - 2\omega \phi'^2+6A'^2-8\phi'A'+\nonumber\\
&&+4\alpha e^{-2B} \left(-3A'^4+24 A'^3\phi'-
36b A'^2\phi'^2+4a\;(4A'\phi'^3+\phi'^4)-6c \phi'^4] \right)=0
\eea
(note the absence of second derivatives). 
Furthermore a linear combination of the remaining two differential equations
can be directly integrated to give
\bea
\label{1integral}
&&e^{4A-2\phi-3B} \left[ e^{2B} (2 [\omega+1]\phi'+A') \right.\nonumber\\
&&\left.- 4\alpha \left(3A'^3+6 A'^2\phi'-
6b [2A'^2\phi'+ A'\phi'^2]+4a [2A'\phi'^2+\phi'^3]
-4c \phi'^3 \right)\right]=-\C \ ,
\eea
where $\C$ is a constant of integration. 
It is interesting to note that 
for the Kaluza-Klein like choice of parameters~\bref{kkparam}, this
equation factorises,
\be
e^{4A-2\phi-B}(2 [\omega+1]\phi'+A')\left(1 -4\alpha e^{-2B}
[3A'^2 + 6\omega A'\phi' +2\omega (2\omega+1)\phi'^2]\right) = -\C \ .
\ee
The integration constant $\C$ will acquire a physical meaning in the
case of 1 dimensional toroidal compactification. The above system of
equations (\ref{55}--\ref{1integral}) are independent and yield the full
set of solutions of eq.~(\ref{action}) for a conformally flat spacetime.

The brane contribution to the action in the Jordan/string frame is
\be
S_b=-\int d^4x \sqrt{-h} \, T(\phi)
\label{brane}
\ee
where $T$ is the brane tension. We can construct a brane world solution
by joining together two spacetimes $\mathcal{M}_R$ and
$\mathcal{M}_L$. If we choose the coordinates so that $z<0$ in
$\mathcal{M}_L$, $z>0$ in $\mathcal{M}_R$ and $z=0$ on the brane, then
the brane junction conditions are
\be
\label{junct1}
\left[e^{-B}(-3A'+2\phi')+4\alpha e^{-3B}\left(A'^3-6A'^2\phi'
+6bA'\phi'^2-{4a\over 3}\phi'^3 \right)\right]_L^R
= M^{-3}e^{2\phi} T(\phi)
\ee
\bea
\label{junct2}
&&\left[ e^{-B}(-4\omega\phi'-8A')+ 4\alpha e^{-3B}\left(8A'^3-24bA'^2\phi'
+16a(A'\phi'^2+\phi'^3/3)-8c\phi'^3 \right) \right]_L^R\nonumber \\
&&= -\,M^{-3}e^{2\phi} \frac{dT(\phi)}{d\phi}
\eea
where $\left[X\right]_L^R=X_{R}-X_{L}$ denotes the jump of X across the brane.
It is interesting to note~\cite{Jakobek} that a linear combination of
the above gives a similar equation to (\ref{1integral}), 
\bea
\label{comb}
\left[e^{-B}(2(\omega+1)\phi'+A')-4\alpha e^{-2B}\left(3(A'^3+2A'^2\phi')
-6b(2A'^2\phi'+A'\phi'^2) 
\right.\right. \nonumber \\ \left.\left.
+4a(2A'\phi'^2+\phi'^3)-4c\phi'^3 \right) \right]_L^R
=\frac{M^{-3}}{2}\frac{d}{d\phi} \left( e^{2\phi}T(\phi)\right) \ .
\eea
Thus if the integration constants in (\ref{1integral}) are such that
$[\C]^R_L=0$ (which corresponds to $\C=0$ in the $Z_2$-symmetric case),
the RHS of (\ref{comb}) must vanish. This constrains the brane tension
to be $T(\phi) \propto e^{-2\phi}$, which is the natural choice for the
dilaton coupling to NS-branes in heterotic string theory. Such a case
was studied in refs.~\cite{mavromatos2, us}. In ref.~\cite{us} in
particular all requirements for `self-tuning' (absence of naked
singularities, finite volume element) were verified apart from
`self-tuning' itself. This was precisely due to the fact that the
integration constant was set to be zero. Generically the $[\C]^R_L=0$
solutions always require fine-tuning of the fundamental parameters. We
will investigate in detail a case where $\C\neq 0$ in section~\ref{KK6D}.

\section{Brane gravity and stability}
 
\label{gm}

In order to investigate gravity as perceived by a four-dimensional observer
and stability issues concerning the background solutions 
we will in this section consider perturbations around the general
background~\bref{ansatzB}. This will permit us to understand some
important characteristics proper to the Einstein-Gauss-Bonnet theory
that we are investigating. In this
paper we will concentrate on the spin 2 perturbations 
which correspond to tensorial  gravity on the four-dimensional
brane world. We will see that the wave equation operator for the
graviton modes in Lovelock gravity is closely analogous to that of
Einstein gravity (see~\cite{deruelle} for a recent discussion and
references within). This
follows from the fact that the two theories were constructed using
similar requirements for their gravitational properties. The junction
conditions will be however crucially different. 

Start by choosing an axial gauge for the perturbed
metric ($\gamma_{a5}=0$): 
\be
ds^2 = e^{2A(z)}(\eta_{\mu\nu}+\gamma_{\mu\nu}(x,z)) dx^\mu dx^\nu 
+ e^{2B(z)} dz^2
\ee
and consider a perturbed scalar field
$\phi(z) + \varphi(x,z)$, with $\gamma_{\mu\nu}$ and $\varphi$
small. The first order perturbation to the $(\mu\nu)$-component of the
gravitational field equation yields a rather complicated expression,
\bea
&&\pa_z (p(z) \pa_z[\gamma_{\mu\nu} - \eta_{\mu\nu}\gamma]) 
- w(z)
\left(2\pa^\sigma\pa_{(\mu} \gamma_{\nu)\sigma} 
- \Box_4 \gamma_{\mu\nu} - \pa_\mu \pa_\nu \gamma 
- \eta_{\mu\nu}[\pa^\sigma \pa^\kappa \gamma_{\sigma \kappa} 
- \Box_4 \gamma]\right)
\nonumber \\
&+& \eta_{\mu\nu} (f_1(z)\pa_z^2 \varphi+f_2(z) \pa_z \varphi)
+f_3(z)[\pa_\mu \pa_\nu \varphi - \eta_{\mu\nu}\Box_4\varphi] \nonumber \\
&=&\delta(z) e^{4A} \left(\frac{dT(\phi)}{d\phi} + 2T(\phi)\right)
\eta_{\mu\nu}\varphi
\label{pertab}
\eea
where indices are raised and lowered with $\eta_{\mu\nu}$,
$\Box_4 = \eta^{\mu\nu}\pa_\mu \pa_\nu$ is the flat
four-dimensional Laplacian and $f_1$, $f_2$ and $f_3$ are some given
background 
functions of $z$ that we will not need for what follows. What
will be important however, are the functions $w$ and $p$ which are given by, 
\be
w(z) = e^{2A-2\phi+B}\left[ 1+4\alpha q(z) \right] +4\alpha v'(z) \ ,
\label{wdef}
\ee
\be
p(z) = e^{4A-2\phi-B}\left(1 - 4\alpha q(z)\right) \ ,
\label{pdef}
\ee
where in turn
\be
q(z)=e^{-2B}(A'^2-4A'\phi'+2b\phi'^2) \ ,
\label{qdef}
\ee
\be
v(z)=-e^{2A-2\phi-B}(A'-2\phi') \ .
\label{vdef}
\ee

We now project out the transverse traceless components of
$\gamma_{\mu\nu}$~\cite{anselmi} (see also~\cite{scagrav}). Define
symbolically the operator
$\pi_{\mu\nu} = \eta_{\mu\nu} - \pa_\mu \pa_\nu/\Box_4$, and consider 
\be
\label{proj}
\bar \gamma_{\mu\nu} = \left(\pi_{\mu(\sigma}\pi_{\kappa)\nu} 
- \frac{1}{3}\pi_{\mu\nu}\pi_{\sigma \kappa}\right) \gamma^{\sigma\kappa} \ .
\ee
This satisfies $\pa^\mu \bar \gamma_{\mu\nu} =0$ and 
$\eta^{\mu\nu}\bar \gamma_{\mu\nu} =0$. Then, applying the
projection~\bref{proj} to the perturbation equation~\bref{pertab}, we
obtain the bulk perturbation equation for the tensor modes
\be
\pa_z (p(z) \pa_z \bar\gamma_{\mu\nu}) 
+ w(z) \Box_4 \bar\gamma_{\mu\nu} = 0 \ .
\label{pertproj}
\ee
Note that the form of the perturbation operator for $\bar
\gamma_{\mu\nu}$ remains unchanged for $\alpha\neq 0$. In particular it
is free from derivatives of higher than second order. What changes
however is the interaction with the background i.e.\ the background
dependent functions $w(z)$ and $p(z)$ which are now of higher order in
derivatives than when $\alpha=0$. This fact is spelt out by the
definitions of $q$ and $v$ which are of first order in derivatives. Note
in particular the presence of second order derivatives, $v'$, in the
definition of $w$. This will yield important differences to the Einstein
case for the perturbed junction conditions on the brane.

Given the spacetime symmetries
consider now a plane wave separation of variables:
\be
\bar \gamma_{\mu\nu}(x,z) = \bar c_{\mu\nu}^m(x)\,\psi_m(z) \hspace*{1 cm}
\mathrm{ with } \hspace*{1 cm} \Box_4 \bar c_{\mu\nu}^m(x) = m^2 \, \bar
c_{\mu\nu}^m(x) 
\ee
where $\bar c_{\mu\nu}^m(x)$ is transverse-traceless and $m^2$ is the
mass squared of the graviton mode as perceived by a
four-dimensional observer.
The scalar interaction wave equation (\ref{pertproj}) now reduces to
\be
\label{wavequa}
\left(p(z) \psi_m' \right)' 
+ m^2 \,w(z)\, \psi_m=0 
\ee
valid everywhere in the bulk.
>From the explicit form of $w$ and $p$ in eqs.~(\ref{wdef}) and (\ref{pdef}),
the associated junction conditions at the
brane are given by 
\be
\label{jcp}
[e^{-B}\psi_{m}' ]_L^R=
4\alpha [q e^{-B} \psi_{m}' +
 e^{-2A-B} m^2 \psi_{m} (A'-2\phi')]_L^R \ .
\ee
Furthermore, in order for the perturbed bulk to
induce a well-defined metric on the brane, the modes $\psi_{m}$ have to
be continuous across the brane.

In order to study the propagation of gravitons (existence of zero mode,
stability, and orthogonality of eigenmodes) modulo the initial
conditions supplied by the brane we now need to examine the boundary
value problem as defined by (\ref{wavequa}) and (\ref{jcp}). Such
problems are generically treated in two ways; either as a
Sturm-Liouville boundary problem, or by coordinate transforming to a
Schr\"{o}dinger type equation. Let us follow these methods in turn
pointing out the essential differences appearing here due to the higher
curvature term corrections.

If $p$ and $w$ are positive, the LHS of (\ref{wavequa}) is the usual
`self-adjoint' type differential operator with weight function $w$
just like in standard (singular) Sturm-Liouville theory. The boundary
conditions for (\ref{wavequa}) are however unusual and call for
particular attention. For massless modes $m^2=0$, and $Z_2$ symmetry,
(\ref{jcp}) tells us that $\psi_{m}'=0$ for $z=0$ just like for the
$\alpha=0$ case -- a standard Neumann boundary condition on the
brane. For the massive modes however this is not the case, we have
rather a mixed boundary condition and furthermore the mass of the mode
in question appears in the boundary condition itself (\ref{jcp}). This
is precisely due to the presence of higher derivatives in $w$ and is
thus a feature of Gauss-Bonnet gravity\footnote{Not surprisingly similar
boundary conditions also appear in the DGP model~\cite{DGP} with the
inclusion of the induced gravity term on the brane.}. We can however
treat this in an analogous way by adding a suitable boundary term
(total derivative) to the usual scalar product of the eigenvector
functional space. Standard Sturm-Liouville considerations will then
follow through. Indeed define
\be
\label{product}
(\phi,\psi)=\int_\mathrm{Bulk}{\phi \psi w}\, dz + 4\alpha [\psi \phi v]_L^R
\ee 
where not only $w$ but also $[v]$ has to satisfy positivity
conditions. We have included a non-trivial boundary term which arises
from the second order derivative appearing in $w$~(\ref{wdef}). Suitable
Sturm-Liouville boundary conditions are imposed at asymptotic
infinity. The problem~(\ref{wavequa}--\ref{jcp}) is now self-adjoint. The
modes are then orthogonal, form a complete basis and the problem is free
of tachyon modes. The particular boundary conditions imposed 
here~(\ref{jcp}) lead us to demand that $[v]$ is also positive along
with $w$ and $p$.

If on the other hand $p$ is negative, the sign of the kinetic term in an
effective action for $\psi_m$ will also be negative. The solution will
then have a graviton ghost in the bulk. If $w$ and $p$ have opposite
signs the corresponding effective mass term will be negative. In this
case the system may have tachyon modes and thus be classically
unstable. Note that a negative value of $[v]$ can also permit tachyon
modes to exist, even if $w>0$ (see section~\ref{KK6D} for an
example). This is due to the mixed boundary conditions~(\ref{jcp}) and
essentially means that brane world backgrounds which are stable in
Einstein gravity can develop an instability once we allow for $\alpha$
corrections. This amounts to a higher order curvature correction
instability. This is the most important consequence of the mixed
boundary conditions (\ref{jcp}) and is independent of the presence of
the scalar field in the background. If $p$ and $w$ (and $[v]$) are
negative the system will be classically stable although a graviton ghost
is problematic at the quantum level. Note that the possibility of
negative $p$ or $w$ (or $[v]$) only occurs when the $\alpha$ dependent
terms are included in the theory. It should also be noted that even if
there are no graviton ghosts or tachyons, there will be analogous
issues with the scalar field perturbations (tachyons, ghosts and also
radiative corrections), although we will not consider them in this
paper.

A necessary condition for gravity to be localised on the brane is that
the zero mode solution of (\ref{wavequa}--\ref{jcp}) be
normalisable. This is equivalent to asking that
\be
\label{local}
\int_{\mathrm{Bulk}} w(z)\, |\psi_0|^2 \, dz < \infty \ .
\ee
The zero mode solution of the bulk equation~\bref{wavequa} is
\be
\label{mozero}
\psi_0=c_0+c_1 \int_\mathrm{Bulk}{dz\over p} \ ,
\ee
where $c_0$ and $c_1$ are constants fixed by the boundary
conditions. For a $Z_2$ symmetric brane world, $c_1=0$.
If the zero mode $\psi_0$ is constant the normalisation
condition~\bref{local} amounts to having a finite four-dimensional volume
element:
\be
\label{Volel}
\int_{\mathrm{Bulk}} w(z)\, dz \ ,
\ee 
which includes the $\alpha$ contributions. If the volume
element~\bref{Volel} is not finite and $p>0$ there will be no
normalisable zero mode. However if we consider spacetimes where $p$ has
a different sign on each side of the brane, this argument may be
bypassed. If $w$ also changes sign the system may be classically
stable, although it will have a ghost on one side of the brane. We
describe an explicit example of this in section~\ref{KK6D}.

It is also useful to recast the wave equation~\bref{wavequa} in a
Schr\"{o}dinger form in the bulk ($z \neq 0$):
\be
\label{schro}
\bigg[ -\frac{d^2}{dx^2}\,+\,V(x) \bigg] \Phi_m = m^2\,\Phi_m
\ee
with $dx/dz=\sqrt{w/p}$ and $\Phi_m = (w p)^{1/4}\,\psi_m$. The effective
potential is then given by,
$$
V(x)= {\frac{d^2}{dx^2}[(wp)^{1/4}]\over(wp)^{1/4}} \ .
$$
Note that when $\alpha \neq 0$ this equation is not valid on the brane
since the coordinate $x$ is ill defined there. The Schr\"{o}dinger picture is
however interesting when studying the large $z$ behaviour. 

\section{Conformally flat solutions}

\label{cfs}

We will now give a method for obtaining analytically the general solution of 
the bulk field equations~(\ref{55}--\ref{1integral}). If $\C=0$ then
$e^{-B} A'$ and $e^{-B} \phi'$ are constants and the differential equations
reduce to algebraic equations, which are easily analysed. Such solutions
have been investigated in~\cite{mavromatos2, Jakobek, us}.
To deal with the more complicated case of $\C \neq$ 0, we begin by choosing our
variable to be $z =d\phi/dA$ (so in these coordinates, the brane is not
necessarily at $z=0$).

The field equations now take the form,
\bea
\label{step1}
{\Lambda} + A'^2 e^{-2B} \left[P_0(z)-4\alpha e^{-2B} A'^2 P_1(z)\right]=0\\
\label{step11} \C e^{2\phi-4A}=
-e^{-B}A'\left[Q_0(z)-4\alpha e^{-2B} A'^2 Q_1(z)\right]
\eea
where
$$
P_0=6-8z-2\omega z^2 \ , \qquad P_1=3-24z+36bz^2- 16az^3-2z^4(2a-3c) \ ,
$$
$$
Q_0=1+2(\omega+1)z \ , \qquad Q_1=3+6(1-2b)z+2z^2(4a-3b)+4z^3(a-c) \ .
$$
For $\Lambda=0$, \bref{step1} is easily solved to give an expression
for $B$
\be
B =\ln A' + \frac{1}{2}\ln \frac{4\alpha P_1}{P_0} \ .
\label{Bz}
\ee
Substituting this into \bref{step11} provides an expression for
$\phi-2A$ in terms of $z$
\be
\phi-2A = \frac{1}{2}\ln (P_0Q_1-Q_0P_1) +\frac{1}{4}\ln\frac{P_0}{P_1^3}
-\frac{1}{2}\ln (2\C\sqrt{\alpha}) \ .
\label{Pz}
\ee
Differentiating with respect to $z$ and rearranging this expression gives
\be
A' = \frac{1}{4(z-2)} \frac{d}{dz}\left\{ 
2\ln(P_0Q_1-Q_0P_1)+ \ln P_0 -3\ln P_1\right\} \ .
\label{Az}
\ee

Integrating this gives $A(z)$, which can then be substituted into
eqs.~\bref{Bz} and \bref{Pz} to find $B(z)$ and $\phi(z)$. It is obvious
that the choice of convenient $a$, $b$ and $c$ can simplify the relevant
algebra. A similar approach can be used when $\Lambda \neq 0$,
although in this case the expression for $B$ \bref{Bz} will be a
solution to the quadratic \bref{step1}, which complicates all
the subsequent steps.

The properties of these solutions can be analysed by expressing the
above polynomials in terms of their roots. Write 
$P_0 = 6(1-u_1z)(1-u_2z)$, $P_1=3\prod_{i=1}^4 (1-v_iz)$, 
$P_0Q_1-Q_0P_1 =15\prod_{i=1}^5 (1-w_iz)$. Some of the $u_i,v_i,w_i$ may
be zero, or complex. For example, if $a=b=c=\omega=0$ all these
parameters are zero except $u_1=4/3$, $v_1=8$, $w_1=-2$. This case will
be discussed in detail in section~\ref{KK6D}.

Assuming that $z=2$ is not a root of any of the three polynomials, the
expressions (\ref{Bz}--\ref{Pz}) imply
\be
A = \frac{1}{4} \sum_i \frac{u_i}{1-2u_i} \ln|1-u_iz|
-\frac{3}{4} \sum_i \frac{v_i}{1-2v_i} \ln|1-v_iz|
+\frac{1}{2} \sum_i \frac{w_i}{1-2w_i} \ln|1-w_iz| + \mathrm{cst.}
\label{Ag}
\ee
\be
\phi = \frac{1}{4} \sum_i \frac{1}{1-2u_i} \ln|1-u_iz|
-\frac{3}{4} \sum_i \frac{1}{1-2v_i} \ln|1-v_iz|
+\frac{1}{2} \sum_i \frac{1}{1-2w_i} \ln|1-w_iz| + \mathrm{cst.}
\label{Pg}
\ee
and
\be
B = \ln A' - \frac{1}{2} \sum_i \ln(1-u_iz) +\frac{1}{2} \sum_i \ln(1-v_iz)
+ \ln\sqrt{2\alpha} \ .
\label{Bg}
\ee
It is clear from the above expressions that the points 
$z = 1/u_i, 1/v_i, 1/w_i$ (real values of $z$ only) are
singularities of some kind. We will refer to these as critical
points. Note that the metric (and $\phi$) are only singular at the point
$z= \pm \infty$ if one of the critical points is zero.

We will assume (for simplicity) that all the roots are distinct. In this
case the behaviour of the metric and the curvature tensors near the
singularities can be easily obtained from the above
expressions~(\ref{Bz}--\ref{Pz}). In the Jordan frame, we see that the
points $z=1/v_i$ are curvature singularities which are a finite proper
distance away from other nearby points (i.e.\ $\int e^B dz$ does not
diverge as $z \to 1/v_i$). On the other hand, the points $z=1/u_i, 1/w_i$
are all at infinite proper distance from other points, and are
not curvature singularities. They therefore correspond to spatial
infinities. Hence all the roots of $P_0$, $P_1$ and $P_0Q_1-Q_0P_1$ are
either naked curvature singularities or spatial infinities. A slightly
different analysis is required for $1/z \to 0$, but the same results
apply. The above expressions actually give several solutions; one for
each range of $z$ which does not contain any coordinate singularities,
and for which $B$~\bref{Bg} is real.

The situation is similar in the Einstein frame. The critical points are 
curvature singularities at finite proper distance if the corresponding
critical point has
$1/2<u_i < 2/3$, $v_i<1/2$, $v_i > 1$ or $w_i > 1/2$. Otherwise they are
spatial infinities. The difference between the two frames arises because
$\phi \to \pm \infty$ at the critical points.

We are interested in regions of spacetime which can be used to construct
brane worlds with localised gravity, but which do not have any problems
with naked singularities. If the zero-mode $\psi_0$ is constant as
usual, the condition~(\ref{local}) for gravity to be localised on the
brane is equivalent to require that the volume element~(\ref{Volel}) is
finite. By substituting the
solution~(\ref{Ag}--\ref{Bg}) into eq.~(\ref{Volel}), we can determine
whether it converges near each of the critical points. If it does then
we can construct a suitable brane world using such a region of
spacetime. The volume element~\bref{Volel} will converge near a critical
point if it satisfies $1/2 < w_i <1$, $v_i>2$, $v_i<1/2$ or 
$1/2 < u_i < 2/3$ as appropriate. Thus in the Jordan frame there are ranges of
$u_i$ and $w_i$ for which regions of spacetime without curvature
singularities and finite volume element can be found. However it should
be noted that if $u_i, w_i>1/2$ (or $v_i<1/2$) then the string coupling,
$e^\phi$, diverges at the corresponding critical point. Furthermore the
time taken (as perceived by an observer on the brane) for a photon to
reach the brane from this point ($\int e^{B-A} dz$), is finite if
$w_i > 1/2$, $w_i < 0$, $v_i > 2$, $v_i<1/2$ or $1/2 < u_i < 2/3$.
Hence the region of strong coupling will be visible from the
brane. In the Einstein frame the conditions for a finite volume element
are the same as those for a naked singularity.

If $\Lambda \neq 0$ a similar approach can be used. Asymptotic behaviour
of the solutions can be obtained near the critical points, and this can
be used to determine if the volume element is finite or if there are
naked singularities. As for $\Lambda = 0$ there are no solutions with
finite volume element and no curvature or string coupling
singularities. Similar ideas have previously been used to analyse
the $a=b=c=-\omega=1$ case~\cite{Jakobek}. 

Hence, for $\C \neq 0$, it does not generally seem to be possible to
obtain brane models with finite volume element and without naked
singularities. This is not true for $\C=0$~\cite{Jakobek,us}, but the
brane tension is then fine-tuned, as explained in section~\ref{fa}.

There are several cases for which the above analysis does not apply. If
the polynomial $q$~\bref{qdef} which appears in the definition of the
volume element~\bref{wdef} shares a root with $P_1$, the volume element
will converge for a wider range of parameters. Similarly if $1+4\alpha q$
shares a root with one of the other polynomials. The situation will
also be changed if any of $P_0$, $P_1$, or $P_0Q_1-Q_0P_1$ share a
root. However these situations will not generally arise without
unnatural fine-tuning of the parameters $\omega$, $a$, $b$ and $c$. An
exception is the point $z=\infty$. For many of the Kaluza-Klein like
choices of parameters~\bref{kkparam} this critical point corresponds to
a root of more than one of the polynomials. However each of these
cases has similar properties to the non-degenerate critical points
discussed above.

\section{Six-dimensional Kaluza-Klein case}

\label{KK6D}

Let us now focus on the case where $\phi$ is a moduli field of a compact
sixth dimension. In the Jordan frame, putting $N=1$ in the
action~\bref{caction} corresponds to $\omega=a=b=c=0$. This is obviously
the simplest case to consider for the action~(\ref{action}). We will now
construct the general solution for these parameters.

\subsection{General bulk solution}

We start by considering the 6-dimensional Gauss-Bonnet anti-de Sitter
black hole solution~\cite{Boulware,Cai}. The solution reads,
\be
\label{blackhole}
ds^2_6 = -V(r) dT^2 + r^2 dx_4^2 + \frac{dr^2}{V(r)}
\ee
where
$V(r) =\kappa+ r^2 (1 \pm U)/(12\alpha)$ with
\be
U = \sqrt{1-\frac{\Lambda}{\Lambda_c} + \frac{\mu}{r^5}}
\ee
Here, $dx_4^2$ is an Euclidean space of constant curvature $\kappa$, 
and $\Lambda_c = -5/(12\alpha)$. Note in passing that
if $\Lambda= \Lambda_c$ the field equations have a degenerate
solution with $e^{-B}A' = 1/\sqrt{12\alpha}$ and $\phi$ completely
arbitrary. This corresponds to the Class I solutions found in ref.~\cite{fax}. 

The integration constant $\mu$ is related to the `AD energy' of the
solution.  Unlike ordinary Einstein gravity ($\alpha=0$) there are two
branches of solutions and both vacua are classically
stable~\cite{teykin} to small perturbations.  The solution corresponding
to the $(-)$  choice of sign has a well defined $\alpha\rightarrow 0$
limit, whereas the second solution exists only if $\alpha\neq 0$. For
lack of a better name we will call the $(-)$ branch the Einstein branch,
and the $(+)$ branch the Gauss-Bonnet branch. Note that 
particular attention has
to be given to the `AD energy' of the Gauss-Bonnet branch as was shown
in ref.~\cite{teykin} (see also \cite{pad} for a Hamiltonian approach). 
We emphasise that the Gauss-Bonnet branch does
not have an equivalent solution in Einstein theory.

Let us focus on the planar case $\kappa=0$. 
For both branches there is the standard curvature 
singularity at $r=0$. If $(1-\Lambda/\Lambda_c)\mu<0$ then there will be
an additional singularity at $r=r_* = (\Lambda/\Lambda_c-1)^{-1/5}\mu^{1/5}$, 
where $U=0$. Only the solution with the lower
choice of sign has an event horizon, and then only if $\Lambda\mu<0$. The
horizon is located at $r = r_h = (\Lambda_c/\Lambda)^{1/5}\mu^{1/5}$. Otherwise
there is no black hole but merely a naked singularity.

A conformally flat, five-dimensional scalar field solution is obtained
from the black hole~\bref{blackhole} by the analytic continuation
$t=i x_4$ and $X_1 \propto iT$, comparison with \bref{kk}, and then 
compactification of the sixth dimension. 
The scalar field solution in brane background (\ref{ansatzB}) reads,
\be
\label{kkA}
A = \ln \frac{r}{r_b} + A_b \ ,
\ee
\be
\label{kkphi}
\phi = - \frac{1}{4}\ln \frac{r^2 (1 \pm U)}{r_b^2 (1 \pm U_b)} + \phi_b
\ ,
\ee
\be
\label{kkB}
B = - \frac{1}{2}\ln r^2 (1 \pm U) + \frac{1}{2}\ln 12\alpha \ ,
\ee
where the parameters $A_b$, $\phi_b$ and $r_b$ are constants which will be
convenient when discussing boundaries, and 
$$
U_b=U(r_b) = \sqrt{1-\frac{\Lambda}{\Lambda_c} + \frac{\mu}{r_b^5}} \, .
$$ 
Note that the parameter $\mu$ in the above solution is proportional to
the constant $\C$ in~\bref{1integral}. 

Taking $A_b=(2/3)\phi_b$ the metric in the Einstein frame reads,
\be
\label{bb}
ds_E^2 = e^{-4\phi/3} ds^2 = \left(\frac{r}{r_b}\right)^{8/3}\!
\left(\frac{1 \pm U}{1 \pm U_b}\right)^{1/3}\! dx^\mu dx^\nu \eta_{\mu\nu}
+ \frac{12 \alpha e^{-4\phi_b/3} dr^2}
{r_b^{2/3} r^{4/3} (1 \pm U)^{2/3} (1\pm U_b)^{1/3}}
\ee

If the original vacuum solution has an event horizon, it will be
transformed into a curvature singularity for the scalar field
system. This is a common phenomenon of conformally flat scalar field
spacetimes obtained in this way. Of course a naked singularity signals
the endpoint of the validity of a particular solution. There are a total of 7
distinct types of spacetimes which can be obtained from the
metric~\bref{blackhole} in this way. They can be classified according to
the choice of sign in the definition of $V(r)$, the sign of $\mu$, and
the value of $\Lambda$. Most of them have naked curvature singularities
which restrict the allowed range of $r$.

For the Gauss-Bonnet branch, if $\Lambda < \Lambda_c$ (in which case
$\mu$ must be positive), the metric is only well-defined in the
coordinate range $0 < r < r_*$. This solution has two naked
singularities which are a finite proper distance apart. If
$\Lambda > \Lambda_c$ the solution is well-defined for $r > r_*$ if $\mu < 0$,
and for $r>0$ if $\mu \geq 0$. These solutions have one naked singularity
and are of infinite proper distance i.e.\ infinite in the extra dimension.

For the Einstein branch we have more possibilities due to the additional
singular point $r=r_h$. The metric is only well-defined for
$r_h < r < r_*$ if $\Lambda < \Lambda_c$ (and $\mu>0$), and for
$r_* < r < r_h$ if $\Lambda > 0$ (in which case $\mu$ must be negative).
Again we have two naked singularities separated by finite proper
distance. When $\Lambda_c < \Lambda < 0$ the solution is well defined
for $r > r_h$ if $\mu>0$ and for $r > r_*$ if $\mu<0$ (or just simply $r>0$ if
$\mu=0$). If $\Lambda=0$ for the Einstein branch, $\mu$ must be negative
and the metric may be defined for $r > r_*$, but has a naked singularity
at infinity ($r \rightarrow \infty$).

Finally for $\Lambda=\Lambda_c$, there is only one branch  {\footnote{In
$D=5$ this case is related to Chern-Simons gravity (see for 
instance~\cite{scan})}}, $1 \pm U$ being replaced by 
$1+\bar \mu/r^{5/2}$. The solution is well-defined for 
$r> (-\bar \mu)^{2/5}$ if $\bar\mu<0$ and for $r>0$ if $\bar \mu>0$.

For comparison with the previous section, the three $\Lambda=0$
solutions: Gauss-Bonnet branch with $\mu<0$, Gauss-Bonnet branch with
$\mu>0$, and Einstein branch with $\mu<0$ respectively correspond to
the three ranges $-\infty < z < -1/2$,
$-1/2 < z < 1/8$ and $3/4 <z < \infty$ of the coordinate
$z=d\phi/dA$. The points $z=-1/2, 3/4$ are spatial infinities
($r=\infty$), while $z=0,1/8$ are the curvature singularities ($r=r_*, 0$
respectively).

An infinite (proper distance) brane world solution can be constructed
by taking two halves of infinite bulk spacetimes (not necessarily
identical) and `gluing' them together. The brane acts as a boundary for
the bulk spacetime on both sides permitting the elimination of one
curvature  singularity. We see immediately that spacetimes with two
naked  singularities (which are also of finite proper distance) are
unsuitable for constructing singularity free brane worlds. Thus we can
construct singularity free brane world models using the any of the
$\Lambda \geq \Lambda_c$ Gauss-Bonnet branch solutions or the 
$\Lambda_c \leq \Lambda < 0$ Einstein branch solutions, provided that we
keep the appropriate singularity free half of the bulk spacetime. Note that we
keep asymptotic infinity $r\rightarrow \infty$ which represents proper
infinity too. We take the brane to be positioned at $r=r_b$. The
constants $A_b$ and $\phi_b$ have to be the same on both sides of the
brane, in order to guarantee a well-defined induced metric as well as
continuity of the scalar field. In contrast, the parameter $\mu$, or
equivalently $U_b$, may take different values on each side.

Before moving on to boundary conditions we note in passing that
if the black hole horizon is not flat then analytic continuation (see
for instance~\cite{padilla} for an $\alpha=0$ example) gives
in the Einstein frame,
\be
\label{ds}
ds_E^2=V^{1/3}r^2(-dt^2+e^{2\sqrt{\kappa}t}dx_3^2)+V^{-2/3}dr^2
\ee
with $\phi=-{1\over 4}\ln V(r)+\phi_0$. Notice then how the curvature of
the horizon after analytic continuation translates into the de Sitter 
or anti-de Sitter 
expansion of the brane world sections. The metric (\ref{ds}) represents the
bulk background for a brane world of constant curvature. 
Having obtained the full set of
spacetime solutions we now can move on to the boundary conditions.

\subsection{Brane tension and parameter tuning}

\label{tension}

For a $Z_2$-symmetric brane world, the solution (\ref{kkA}--\ref{kkB})
is defined up to 2 integration constants, namely $\phi_b$ and $\mu$ (the
constant $A_b$ is not physical and may be taken to vanish), and 2
possible branches. Applying the junction conditions (\ref{junct1}) and
(\ref{junct2}) to this solution, and taking the brane tension to be 
$T(\phi) = T_1 e^{-\chi \phi}$, we find
\be
\label{chi}
\chi=2+\frac{30\mu}{(U_b^2 \mp 16U_b-32+15U_0^2) r_b^5}
\ee
and
\be
\label{kkT}
T_1 = -\,\frac{M^3}{\sqrt{12\alpha}}\,\frac{32 \pm 16U_b-U_b^2-15U_0^2}{6(1
\pm U_b)^{1/2}} \,e^{(\chi-2)\phi_b} 
= \frac{5\mu\;M^{3}}{\sqrt{12\alpha} r_b^5 (\chi-2) \sqrt{1\pm U_b}} 
e^{(\chi-2)\phi_b} 
\ee

A few comments are now in order. If $\mu=0$ then $\chi=2$ and therefore
$T_1$ must be fine-tuned~\cite{us}. Otherwise fixing $\chi$, the dilaton
coupling to the brane, fixes the value of the integration constant $\mu$
for given $\Lambda$. Then, the brane tension parameter $T_1$ in eq.~(\ref{kkT})
still depends on the arbitrary integration constant $\phi_b$ in
eq.~(\ref{kkphi}), which could conceivably tune itself to allow
four-dimensional flat solutions for any values of the tension. This is
the basic idea behind `self-tuning'. It is also easy to see that the
brane tension can be positive for both branches.

If we relax the assumption of $Z_2$-symmetry across the brane, the
parameter $\mu$ may take different values on each side of the brane, and
so we will have one extra arbitrary integration constant in the
expressions for $\chi$ and $T_1$.

\subsection{Stability and four-dimensional effective gravity}

The functions appearing in the graviton
equations~(\ref{wdef}--\ref{vdef}) are:
\be
\label{wkk1}
w = \mp \sqrt{3\alpha} \frac{r^2}{r_b^3} 
\frac{50U^2U_0^2-U^4-25U_0^4}{12U^3\sqrt{1 \pm U_b}}\,e^{2A_b-2\phi_b}
\ee
\be
\label{pkk1}
p = \mp\frac{r^6}{r_b^5}\frac{(1 \pm U)(U^2+5U_0^2)}{12U\sqrt{3\alpha}
\sqrt{1 \pm U_b}}\,e^{4A_b-2\phi_b}
\ee
\be
\label{vkk1}
v = \mp \frac{r^3}{r_b^3}
\frac{3U^2 \pm 8U+5U_0^2}{8\sqrt{3\alpha} U \sqrt{1 \pm U_b}}\,e^{2A_b-2\phi_b}
\ee
where $U_0 = U(\mu=0)= \sqrt{1-\Lambda/\Lambda_c}$ and the lower sign
corresponds to the Einstein branch. In these coordinates
$r$ ranges from $r_b$ to $\infty$.  We will only consider the cases
without naked bulk singularities ($U_0 \geq 0$ for the Gauss-Bonnet
branch, and $0 \leq U_0 < 1$ for the Einstein branch).

First note 
 that the bulk graviton equation~\bref{wavequa} is singular in the
limit  $r \to \infty$, since the functions $w$ (\ref{wkk1}) and $p$
(\ref{pkk1}) blow up. This is always true except for the trivial case of the
 constant zero mode. In the  Schr\"{o}dinger formulation \bref{schro},
this translates  into a potential $V(x)$ diverging at a finite distance
from the brane in the $x$-coordinate. The situation is thus similar to
that of a particle confined in a box, and we shall impose that the
wave-functions $\Phi_m$ vanish ``at the edge of the box''. This is
equivalent to require:
\be
\label{bound} 
\psi_m \sim \mathit{o}\left(\frac{1}{r^2}\right) \hspace*{0.7 cm} \mathrm{for}
\hspace*{0.7 cm} r\rightarrow \infty 
\ee
All the modes satisfying eq.~(\ref{bound}) are normalisable on the bulk
interval with respect to the weighting function  $w(r)$~(\ref{wkk1}).
Considering the asymptotic behaviour of the wave equation~\bref{wavequa}
with the expressions \bref{wkk1} and \bref{pkk1} shows that
eq.~\bref{bound} selects a one-parameter family of solutions for
$\psi_m$ on each side of the brane. Continuity on the brane and the
junction conditions~\bref{jcp} then result in a homogeneous system of
two equations for the two unknown parameters. Vanishing of its
determinant then gives the allowed values for $m^2$, which are
therefore discrete. 
Since $\alpha$ is the only other mass scale appearing in
the problem, we expect the order of magnitude of
mass spacings between the modes to be:
\be
\label{gap}
\Delta m \sim \frac{1}{\sqrt{\alpha}}
\ee

Secondly note from eq.~\bref{pkk1} that the sign of $p$ is always negative for
the Gauss-Bonnet branch. This implies that the sign of the kinetic term
in an effective action for $\bar \gamma_{\mu\nu}$ is also negative. This
branch therefore suffers for a graviton ghost in the bulk. The same is
true for the corresponding original six-dimensional black-hole
vacuum~\cite{Boulware}, although it has been argued to be classically
stable~\cite{teykin}. On the other hand, $p$ is always positive for the
Einstein branch, so this case is ghost free. As discussed in
section~\ref{gm}, there may be tachyon modes if either $w$ or $[v]$ is
negative. This could occur for the Einstein branch, in which case the
solution would be unstable. 

To illustrate the above two points consider the case $\mu=0$. We can
then find analytic expressions for the modes. Indeed define:
\be
\lambda^2_\pm=|m^2| \, \frac{12 \alpha}{1\pm U_0} r_b^2 \, e^{-2A_b} \ . 
\ee
For $m^2>0$ the bulk solution satisfying \bref{bound} is
\be
\label{states}
\psi_{m}\propto  \,\frac{1}{r^{5/2}}\,
J_{5/2}\left(\frac{\lambda_\pm}{r}\right)
\ee
and the mass spacings between the Kaluza-Klein states are given by:
\be
\Delta m^2 \sim \frac{1-U_0}{12\alpha}\,e^{A_b} 
\ee
in agreement with (\ref{gap}).\\
Now consider 
a solution with $m^2 < 0$, satisfying (\ref{bound}) at infinity, 
\be
\label{tachyon}
\psi_{m} \propto \frac{1}{r^{5/2}} \, 
I_{5/2}\left(\frac{\lambda_{\pm}}{r}\right) \ . 
\ee
In the case of two Einstein branches on each side of the brane, the
corresponding junction condition~\bref{jcp} then implies
\be
\frac{\lambda_-}{r_b} \, I_{5/2}\left(\frac{\lambda_-}{r}\right)
=\frac{3U_0}{4(1-U_0)}\, I_{3/2}\left(\frac{\lambda_-}{r}\right)
\ee
If $0<U_0<1$, this equation is always satisfied by one real value of
$\lambda_-$, and so there is always a tachyon mode when $\mu=0$ for the
Einstein branch, even though $p$ and $w$ are positive. For small $U_0$,
$m^2 \sim -(U_0/\alpha) e^{2A_b}/r_b^2$. The higher order curvature term
has destabilised the solution. Note in contrast that 
the junction conditions do not allow for any tachyonic gravitons in the
case where the Gauss-Bonnet branch is present on at least one side of
the brane (and $\mu=0$). 

To ensure that $w>0$ we require
$5(5-2\sqrt{6}) U_0^2 < U_b^2 < 5(5+2\sqrt{6}) U_0^2$. To ensure $[v] \geq 0$ 
we need $U_b \leq (4/3)(1-\sqrt{1-15U_0^2/16})$, and so $\mu$ must be
negative. It is only possible to satisfy both these conditions
simultaneously if $U_0 \gtrsim 0.87$, hence $\Lambda/\Lambda_c$ must be
small if the spacetime is to be free from tachyons and ghosts.

As for the constant zero mode, we see from eq.~(\ref{wkk1}) that the volume
element~(\ref{Volel}) is infinite for the
solution~(\ref{kkA}--\ref{kkB}), in agreement with the general
discussion of section~\ref{cfs}. In the present example this is
because, in order to avoid naked singularities, we have kept the region
of space-time which contains the conformal infinity of the original
six-dimensional space~\bref{blackhole}, which is asymptotically anti-de
Sitter. Nevertheless, it is possible to localise gravity on the brane, 
if we break $Z_2$ symmetry by joining together an
Einstein with a Gauss-Bonnet branch. The price to pay  
however is a bulk graviton ghost on the Gauss-Bonnet side. 
Indeed, for $m^2=0$, the solution
(\ref{mozero}) of the  perturbation equation (\ref{wavequa}) which
satisfies the boundary condition~(\ref{bound}) is given by:
\be
\label{zeromo}
\psi_0= C\;\left[\pm \frac{5U_0^2}{2} 
\ln \left(\frac{U^2}{6U_0^2}+\frac{5}{6}\right) 
\pm \ln\left(\frac{1\pm U}{1\pm U_0}\right)
- \sqrt{5}U_0 \, \arctan \frac{\sqrt{5}(U-U_0)}{U+5U_0}\right]
\ee
for $\mu \neq 0$ and $C/r^5$ if $\mu=0$, where the constant $C$
may be different on both sides of the brane. The junction 
condition~\bref{jcp}, together with continuity of $\psi_0$ on the brane, then
require $\mu \neq 0$ and two different branches of the solution on 
each side of the brane. The special junction conditions for higher
order terms discussed in section~\ref{gm}, as well as the two possible
signs of $p$ (and hence the ghost) are essential here. In this case, the
zero-mode (\ref{zeromo}) is normalisable, and separated from the massive
ones by the mass gap (\ref{gap}). Since it is natural for $\alpha$ to be
of the order of the fundamental scale of the underlying theory, the
contribution of the massive gravitons to brane gravity at low energies
is likely to be negligible. This setup is still free of singularities in
the bulk, and the tension of the brane is not fine-tuned.

\section{Conclusions}

In this paper we have investigated the properties of a Poincar\'e brane
world, located in a five-dimensional background containing a scalar
field. We considered the most general quartic order action which
produced field equations that were linear in second order derivatives,
and contained no higher derivatives. If the scalar field plays the role
of the dilaton, the couplings of the dilatonic higher order terms in the
Lagrangian~(\ref{action}) are constrained in order to reproduce the
scattering string amplitudes on shell~\cite{Mets}. We have also derived their
values (\ref{kaka}) when the scalar field arises as the unique modulus
coming from the Kaluza-Klein toroidal compactification of $N$ extra dimensions.

In order to study how gravity may be perceived by a four-dimensional
observer, we considered background perturbations corresponding to
graviton modes which are tangent to the brane. Although the tensorial
structure of bulk gravitons is still the same as in Einstein gravity,
the higher order terms lead to significant modifications to the original
mechanism of gravity localisation. In particular, they lead to mixed
rather than Neumann boundary conditions (\ref{jcp}), 
much like induced gravity
models~\cite{DGP}, where the momenta of modes appear in the boundary
conditions themselves. This renders questions of eigenmode orthogonality
and of background stability far more subtle and constrained than the
Einstein case (\ref{product}). 
For example it is possible to have a scalar field background which is
stable in Einstein gravity, but which develops tachyon modes when a
Gauss-Bonnet correction  is included. A solution with this type of instability
(\ref{tachyon}) was described in the previous section.

Furthermore, for some non $Z_2$-symmetric configurations,
gravity may be localised on the brane, even though the volume element is
infinite. This involves a non-constant graviton zero mode, and is only
possible if there are multiple branches of solutions, as
occurs in Einstein-Gauss-Bonnet gravity. However, combining
different branches in this way generically implies a bulk graviton ghost
on one side of the brane.

In section~\ref{cfs} we presented a method for analytically obtaining
the general solution of the field equations,  which as we showed in
section~\ref{fa}, is a necessary step if one is to address the
`self-tuning' mechanism in this context (i.e.\ trying to resolve the
four-dimensional cosmological constant problem by allowing
four-dimensional Poincar\'e invariance on the brane to be obtained
without fine-tuning of its tension, due to its conformal coupling to
the scalar field~\cite{ADKSS}). We then argued that requiring finite
volume element, as well as no naked singularities (and finite string
coupling), generally implies fine-tuning rather than self-tuning of
the brane tension.

Finally, to illustrate all of the above, we considered in detail the case where
the scalar field is the modulus of a sixth flat dimension. The
general solution then is simply obtained by analytic continuation of the
corresponding six-dimensional Gauss-Bonnet
black-hole~\cite{Boulware,Cai}. The spatial topology of the black hole
horizon transforms into the four-dimensional spacetime curvature of the
brane thus providing the de Sitter, Poincar\'e, and anti-de Sitter brane
solutions. Regular brane solutions can then be constructed without
fine-tuning of tension, or naked singularities in the bulk. It is rather
intriguing that the zero-mode graviton is generically non-normalisable
i.e.\ gravity is not localised, except if we consider different branches
of solutions on each side of the brane. Such a setup allows for a
normalisable 
zero-mode graviton to be separated from the (discrete) massive ones by a
mass gap, of the order of the fundamental scale coupling of the
higher order terms. The
drawback of this case is the presence of a bulk graviton ghost in the
Gauss-Bonnet branch which signals a potential quantum inconsistency of
the gravitational background. This is a rather subtle problem since the
corresponding original six-dimensional black-hole anti-de Sitter
vacuum~\cite{Boulware} has also been argued to be classically
stable~\cite{teykin} to small perturbations due to the change of sign of
the AD energy in the Gauss-Bonnet branch. The presence, significance and
eventual cure of such bulk ghosts certainly demands further study.

Furthermore, on a different note, higher dimensional toroidal
compactification should relate the scalar field solutions discussed here
(for $N>2$) to the Gauss-Bonnet equivalent of cosmic p-brane
solutions~\cite{ruth}. Such solutions yield backgrounds of higher
co-dimension brane worlds~\cite{roberto,ulrich}. It would be intriguing
to know their Gauss-Bonnet versions and to see if they can solve some of
the problems of their Einstein counterparts such as the presence of the
bulk naked singularity or the fate of self-tuning in higher co-dimension
spacetimes (see for example \cite{nav}).

\subsection*{Acknowledgements}

It is a great pleasure to thank Pierre Binetruy, Brandon Carter, 
Jihad Mourad, Luigi Pilo, Valery Rubakov and Dani Steer 
for encouraging and
discouraging remarks and observations. CC thanks G. Kofinas for discussions on
Chern-Simmons gravity. SCD is grateful to the EC network
HPRN--CT--2000--00152 for financial support.

\appendix
\section{Appendix: Field equations}

Varying the action (\ref{action}) with respect to the metric gives:

\begin{eqnarray}
\label{emunu}
&& f\Lambda g_{ab}
+ fG_{ab} - f'[\na_{\! a}\na_{\! b}\phi - g_{ab}\Box\phi]
- f'' [\na_{\! b}\phi\na_{\! a}\phi - g_{ab} (\na\phi)^2]
\nn \\ &&
{}-4\omega f\left(\na_{\! b}\phi\na_{\! a}\phi
-\frac12 g_{ab} (\na \phi)^2 \right)
+2\alpha \left(fH_{ab} +
 2 P_{aebc}(f''\na^e\phi\na^c\phi +f' \na^e\na^c\phi)\right) 
\nn \\ && 
{}-8b \alpha \left(f[2\na^e\phi\na^c\phi P_{aebc}
+(\na\phi)^2 G_{ab}
+2(\na_{\! a}\na_{\! e}\phi)(\na_{\! b}\na^e\phi)
-2(\na_{\! a}\na_{\! b}\phi)\Box\phi
\right.\nn \\ && \hspace{1.5cm} \left.
{}+(\Box\phi)^2 g_{ab} 
-(\na^e\na^c\phi)(\na_{\! e}\na_{\! c}\phi) g_{ab}]
+f'[(\na \phi)^2\Box\phi g_{ab} 
-(\na^e\na^c\phi)\na_{\! e}\phi\na_{\! c}\phi g_{ab} 
\right.\nn\\ && \hspace{1.5cm} \left.
{}-\na_{\! a}\phi\na_{\! b}\phi\Box\phi 
-(\na \phi)^2\na_{\! a}\na_{\! b}\phi
+2\na^e\phi\na_{(a}\phi\na_{\! b)}\na_{\! e} \phi]\right) 
\nn\\ &&
{}+16a \alpha \left(f[
(\na^e\phi\na^c\phi(\na_{\! e}\na_{\! c}\phi)g_{ab} 
-2\na^e\phi\na_{(a}\phi (\na_{\! b)}\na_{\! e}\phi) 
+\na_{\! a}\phi \na_{\! b}\phi\Box\phi]
\right.\nn\\ && \hspace{1.5cm} \left.
{}+\frac{1}{2}f'[(\na\phi)^2 g_{ab}
-2(\na \phi)^2 \na_{\! a}\phi\na_{\! b}\phi] \right)
\nn\\ &&
{}-16c \alpha f\left(2(\na \phi)^2\na_{\! a}\phi\na_{\! b}\phi
-\frac{1}{2}(\na \phi)^4 g_{ab}\right)
= 0
\end{eqnarray}
where the prime denotes differentiation with respect to $\phi$, and
\be
H_{ab} = 
R_{adec}R_b{}^{dec} -2R^{ec}R_{eacb}
-2R_{ae}R^e{}_b +RR_{ab}
-\frac14 g_{ab}\LGB
\ee
\be
P_{aebc} = 
R_{aebc} + R_{be} g_{ac} + R_{ac} g_{be} 
- R_{ec}g_{ab} - R_{ab} g_{ec} 
+ \frac{1}{2}Rg_{ab} g_{ec} - \frac{1}{2}R g_{be} g_{ac} \ .
\ee

Varying (\ref{action}) with respect to the scalar field gives:
\bea
\label{scal}
&& 0=-8\omega f\Box\phi-4 \omega f'(\nabla \phi)^2
-f'R + 2 f'\Lambda + 2\Lambda' f
+\alpha\left(-f'\LGB-16(3 cf'+af'') \; (\nabla \phi)^4
\right.\nn \\ && \left.{}
-16b \; G^{a b} \, 
(f'\nabla_{\! a}\phi\nabla_{\! b}\phi + 
2f \nabla_{\! a} \nabla_{\! b}\phi)
+32 a f\; (\Box\phi)^2 
-64 c f\; (\nabla \phi)^2 \Box\phi
\right. \nn\\ &&\left.{}
-64 (2cf + af')\; \nabla^{b}\phi \nabla^{a}\phi
\nabla_{\! b} \nabla_{\! a}\phi
-32 a f\; (\nabla^{a} \nabla^{b}\phi \nabla_{\! a} \nabla_{\! b}\phi
+ R^{a b} \nabla_{\! a}\phi\nabla_{\! b}\phi)\right)
\eea

For a conformally flat ansatz of the form~\bref{ansatzB}, the
$(\mu,\nu)$ and (5,5) components of (\ref{emunu}) read, respectively:
\bea
\label{munu}
&&4\phi''-6A''-4(\omega+2)\phi'^2-12A'^2-2\Lambda
e^{2B}+12A'\phi'+6A'B'-4\phi'B'
\nonumber \\ 
&&+ 24\alpha e^{-2B}\left( 
A'^2A''-2A'^2\phi''-4\phi'A'A''+4A'^2\phi'^2-6\phi'A'^3+A'^4-
A'^3B'+6A'^2B'\phi' \right) \nonumber \\ 
&&+48b\;\alpha e^{-2B} \left(\phi'^2A''+2A'\phi'\phi''+2A'^2\phi'^2
-2A'\phi'^3-3A'B'\phi'^2 \right) \nonumber \\ 
&&+32a\alpha e^{-2B}\left(-\phi'^2\phi''+\phi'^4+B'\phi'^3\right)
-16c\alpha e^{-2B}\phi'^4 =0
\\ \nonumber \\ 
\label{5,5} 
&&e^{2B}2\Lambda-4\omega\phi'^2+12A'^2-16\phi'A'
 \nonumber \\ &&{}
+\alpha e^{-2B} \left( 24[-A'^4+8A'^3\phi'] -
288b\;[A'^2\phi'^2]+32a\;[4A'\phi'^3+\phi'^4]-48c\;[\phi'^4] \right)=0
\eea
while the scalar field equation (\ref{scal}) gives:
\bea
\label{scalal} 
&&-4\omega\phi''-8A''+4\omega\phi'^2-20A'^2-2\Lambda e^{2B}
-16\omega A'\phi'+8A'B'+4\omega\phi'B' \nonumber \\ 
&&+\alpha e^{-2B}\left( 24[4A'^2A''+5A'^4-4A'^3B']
-96b\;[A'^2\phi''+2\phi'A'A''-A'^2\phi'^2+4\phi'A'^3-3A'^2\phi'B'] 
\right) \nonumber \\ 
&&+ 32a\;\alpha e^{-2B}\left( 
2\phi'^2A''+2\phi'^2\phi''+4A'\phi'\phi''+8A'^2\phi'^2-\phi'^4
-6A'B'\phi'^2-2B'\phi'^3 \right) \nonumber \\
&&-16c\;\alpha e^{-2B}\left(6\phi'^2\phi''+8A'\phi'^3-3\phi'^4
-6B'\phi'^3 \right)=0
\eea
where a prime denotes derivative with respect to $z$.

\end{document}